\documentclass[12pt,preprint]{aastex}
\usepackage{epsfig}
\usepackage{natbib}
\usepackage{amssymb, amsmath, amsbsy, epsfig, epsf, slashbox, rotating}
\usepackage[dvips]{color}

\newcommand{\A}{$\rm \AA$}

\newcommand{\feii}{Fe\,{\footnotesize II}}

\newcommand{\mgii}{Mg\,{\footnotesize II}}

\newcommand{\civ}{C\,{\footnotesize IV}}

\newcommand{\etal}{et al.}

\shorttitle{Outflow and Hot Dust}\shortauthors{Wang, Huiyuan et al.} 

\begin{document}
\title {Outflow and hot dust emission in high redshift quasars}
\author{Huiyuan Wang\altaffilmark{1}, Feijun Xing\altaffilmark{1}, Kai Zhang\altaffilmark{2}, Tinggui Wang\altaffilmark{1}, Hongyan Zhou\altaffilmark{1,3} and Shaohua Zhang\altaffilmark{3}}
\altaffiltext{1}{Key Laboratory for Research in Galaxies and
Cosmology, Department of Astronomy, University of Science and Technology of China, Chinese
Academy of Sciences, Hefei, Anhui 230026, China; whywang@mail.ustc.edu.cn}
\altaffiltext{2}{Key Laboratory for Research in Galaxies and Cosmology, Shanghai Astronomical Observatory,
Chinese Academy of Sciences, 80 Nandan Road, Shanghai 200030, China}
\altaffiltext{3}{Polar Research Institute of China, Jinqiao Road 451, Shanghai 200136, China}
\begin{abstract}
Correlations of hot dust emission with outflow properties are investigated, based on a large z$\sim2$ non-broad absorption lines quasar sample built from the Wide-field Infrared Survey and the Sloan Digital Sky Survey data releases. We use the near infrared slope and the infrared to UV luminosity ratio to indicate the hot dust emission relative to the emission from the accretion disk. In our luminous quasars, these hot dust emission indicators are almost independent of the fundamental parameters, such as luminosity, Eddington ratio and black hole mass, but moderately dependent on the blueshift and asymmetry index (BAI) and full width at half-maximum (FWHM) of \civ\ lines. Interestingly, the latter two correlations dramatically strengthen with increasing Eddington ratio. We suggest that, in high Eddington ratio quasars, \civ\ regions are dominated by outflows so the BAI and FWHM(\civ) can reliably reflect the general properties and velocity of outflows, respectively. While in low Eddington ratio quasars, \civ\ lines are primarily emitted by virialized gas so the BAI and FWHM(\civ) become less sensitive to outflows. Therefore, the correlations for the highest Eddington ratio quasars are more likely to represent the true dependence of hot dust emission on outflows and the correlations for the entire sample are significantly diluted by the low Eddington ratio quasars. Our results show that an outflow with a large BAI or velocity can double the hot dust emission on average. We suggest that outflows either contain hot dust in themselves or interact with the dusty interstellar medium or torus.
\end{abstract}


\section{Introduction}

Outflows are a common phenomenon and have been thought to be an essential component in the
overall structure of quasars. They leave prominent imprints in quasar
spectra, such as broad absorption lines (BALs; Weymann \etal\ 1991) and blueshifted broad
emission lines (BELs; Gaskell 1982), with a typical velocity of 10000km/s. The blueshift of \civ\
BELs strongly increases with Eddington ratio (Wang et al. 2011, hereafter Wang11; Marziani \& Sulentic 2012) and decreases
with the strength of intrinsic X-ray emission relative to UV (Leighly \& Moore 2004; Richards \etal\ 2011).
Similar but weaker trends are also detected by using parameters measured from BALs (Ganguly \etal\ 2007;
Fan \etal\ 2009; Zhang \etal\ 2010). These studies offer strong support to the scenario that the outflow is regulated by the accretion process (see also Sulentic \etal\ 2000). Moreover, outflows are considered to be able to affect star formation in the host galaxies (Silk \& Rees 1998). Recently, Wang \etal\ (2012) found that the outflow property is correlated with gas metallicity, indicative of a complex interaction between black hole and circum-nuclear star formation. Investigations into outflows therefore provide a powerful way to understand the quasar central engine and co-evolution with host galaxies.

One of the noticeable phenomena associated with outflows is that BAL quasars, especially low-ionization BAL quasars, are on average redder than non-BAL quasars in the UV band (e.g. Weymann \etal\ 1991; Brotherton \etal\ 2001). Lots of works interpreted it as being a result of dust extinction (Voit \etal\ 1993; Reichard \etal\ 2003; Hewett \& Foltz 2003; Dai \etal\ 2008; Jiang \etal\ 2013). Interestingly, an `opposite' trend for BEL outflows, i.e. quasars with large \civ\ blueshift is on average bluer than those with small blueshift, was reported (Reichard \etal\ 2003). BAL and BEL outflows are very likely to represent the same physical component, but viewed from different inclination angles, since they manifest quite similar physical properties and correlations (see Wang11 for a comprehensive discussion). This seeming contradiction can be reconciled by a special geometrical configuration in which dust is associated with outflows. In this case, dust reddening is preferably observed in the outflow direction (Zhang \etal\ 2010). Theoretically, dust absorption could be an important acceleration mechanism for dusty outflows (Scoville \& Norman 1995), besides resonant line absorption (e.g. Murray \etal\ 1995) and magnetocentrifugal acceleration (e.g. Proga 2000; Yuan \etal\ 2012). Indeed, there is an observational indication of dust acceleration in that low-ionization BAL quasars are generally redder than high-ionization BAL quasars.

All of these point toward an important link between outflows and dust. Unfortunately, such a link is under-explored due to lack of large and appropriate samples. Until now, most of observational works just focused on showing the presence of dust on the direction of BAL outflows\footnote{Recently, several works found interesting correlations between the outflow in narrow line region (NLR) and the mid-infrared emission (Zhang \etal\ 2013; H\"{o}nig \etal\ 2013). The outflows discussed here have much higher velocities than those in NLR. Thus we are studying totally different outflow component from those works. }. The Wide-field Infrared Survey Explorer (WISE; Wright \etal\ 2010) surveyed the whole sky at much better sensitivities than previous all sky surveys. In this letter, we take advantage of the WISE and the Sloan Digital Sky Survey (SDSS) data releases to construct a large high-redshift quasar sample. This sample of quasars have well-measured UV line parameters and near-infrared (NIR) magnitudes in their rest frame, and thus are appropriate for our study.

This letter is organized as follows. Sample selection and data reduction are described in Section \ref{secsam}.
We present the correlation analysis in Section \ref{secres}. Finally, we summarize and discuss the results in Section \ref{secsum}. Throughout this letter, we adopt the cold dark matter `concordance'  cosmology with $\rm H0=70 kms^{-1}Mpc^{-1}$, $\Omega_{\rm m}=0.3$, and $\Omega_{\Lambda}=0.7$.

\section{Sample Selection and Near-Infrared Spectral Index}\label{secsam}

From the Fifth Data Release (DR5) of SDSS spectroscopic database
(Schneider \etal\ 2007), we select 4963 quasars in the redshift range of $1.7<z<2.2$,  with
reliable measurements of \civ\ and \mgii\ BELs. The sample is described in detail in Wang11.
Here we only give a brief description and refer the interested reader to Wang11 for more details.
To measure the \civ\ lines, we first fit the local continuum
with a power law in two windows near 1450\A\ and 1700\A, where the contamination from emission lines is weak or absent.
Then we subtract the fitting continuum and use two Gaussians to fit the residual spectra around \civ.
Following Wang11, we adopt the blueshift and asymmetry index (BAI) of \civ\ lines which is the flux ratio of
the blue part to the total profile. The blue part is the part of the \civ\ line at wavelengths shorter than 1549.06\A.
Generally, large BAI quasars have large \civ\ blueshift and significant excess flux in the blue wing of \civ\ BELs (figures 4 and 9 in Wang11). Thus the BAI represents a combination of both outflow velocity and the fraction of lines in the blueshifted component, and indeed characterizes the general properties of the outflow. To reduce the possible uncertainty due to measurement error or bias in redshifts, we adopt the redshifts provided by Hewett \& Wild (2010). We discard all known BAL quasars (Scaringi et al. 2009), because the \civ\ line parameters of these quasars can not be reliably measured.

The \mgii\ BELs are fitted using the same procedures as in Wang \etal\ (2009).
We model the continuum in the spectral regimes near \mgii\ with two components: a power law representing
the featureless continuum and the UV \feii\ templates. After subtracting the model continuum, the broad components of
the \mgii\ doublet lines are each modeled with a truncated five-parameter Gauss-Hermite series profile and the narrow component of each line is modeled with a single Gaussian. Based on the FWHM of \mgii\ BELs, we derive the
black hole masses, $M_{\rm BH}$, using the formula of Wang \etal\ (2009). We then compute the Eddington ratio, $l_{\rm E}=L_{\rm bol}/L_{\rm Edd}$, assuming a constant bolometric correction, $L_{\rm bol}=5.9\lambda L_{\lambda}$(3000\A) (McLure \& Dunlop 2004).

We use a match radius of 3 arcsec to cross-correlate our $z\sim2$ catalog with the WISE All Sky Data Release catalog.
Almost all quasars in our sample are detected by WISE (4858 out of 4963). WISE surveyed the whole sky at four infrared
bands with effective wavelengths of $\lambda_1=$3.35, $\lambda_2=$4.60, $\lambda_3=$11.56, and $\lambda_4=$22.08 $\mu\rm m$, which are referred to as W1, W2, W3 and W4, respectively. There are 4731 quasars with signal to noise ratio ($S/N$) larger than 3 in the first three bands and 2337 quasars with $S/N>3$ in all WISE bands. Since we need a sufficient number of quasars and are only interested in the rest-frame NIR spectral energy distribution (SED), we only use the three short-wavelength bands in the following investigation. Our final working sample thus contains 4731 quasars, which correspond to about 97.4\% of SDSS type-I non-BAL quasars detected by WISE.

The magnitudes at W1, W2 and W3 are converted to $\lambda F_{\lambda}$ according to the published WISE zero points. We then derive the monochromatic luminosities, $\lambda L_{\lambda}$, at the three bands, marked as L$_1$, L$_2$ and L$_3$, respectively. The redshifts of our selected quasars are close to 2, so the rest-frame wavelengths of L$_1$, L$_2$ and L$_3$ are around 1.12, 1.53 and 3.85 $\mu\rm m$, which all move into the NIR band. One useful quantity is the spectral slope of the NIR SED. For a quasar at redshift $z$, we calculate the rest wavelength of L$_i$ as $\lambda^{\rm r}_i=\lambda_i/(1+z)$ ($i=1,2,3$).
Adopting a power law ($\propto\lambda^{\beta_{\rm NIR}}$), we then fit the SED L$_i(\lambda^{\rm r}_i)$ to constrain the slope ($\beta_{\rm NIR}$) for this quasar. According to the slope distribution shown in Figure \ref{fig1} and the color correction factors published in Wright \etal\ (2010), the color correction for our quasars is tiny. We thus don't apply any color correction to the WISE fluxes.

\section{Correlation Analysis}\label{secres}

\begin{figure}
\epsscale{1}\plotone{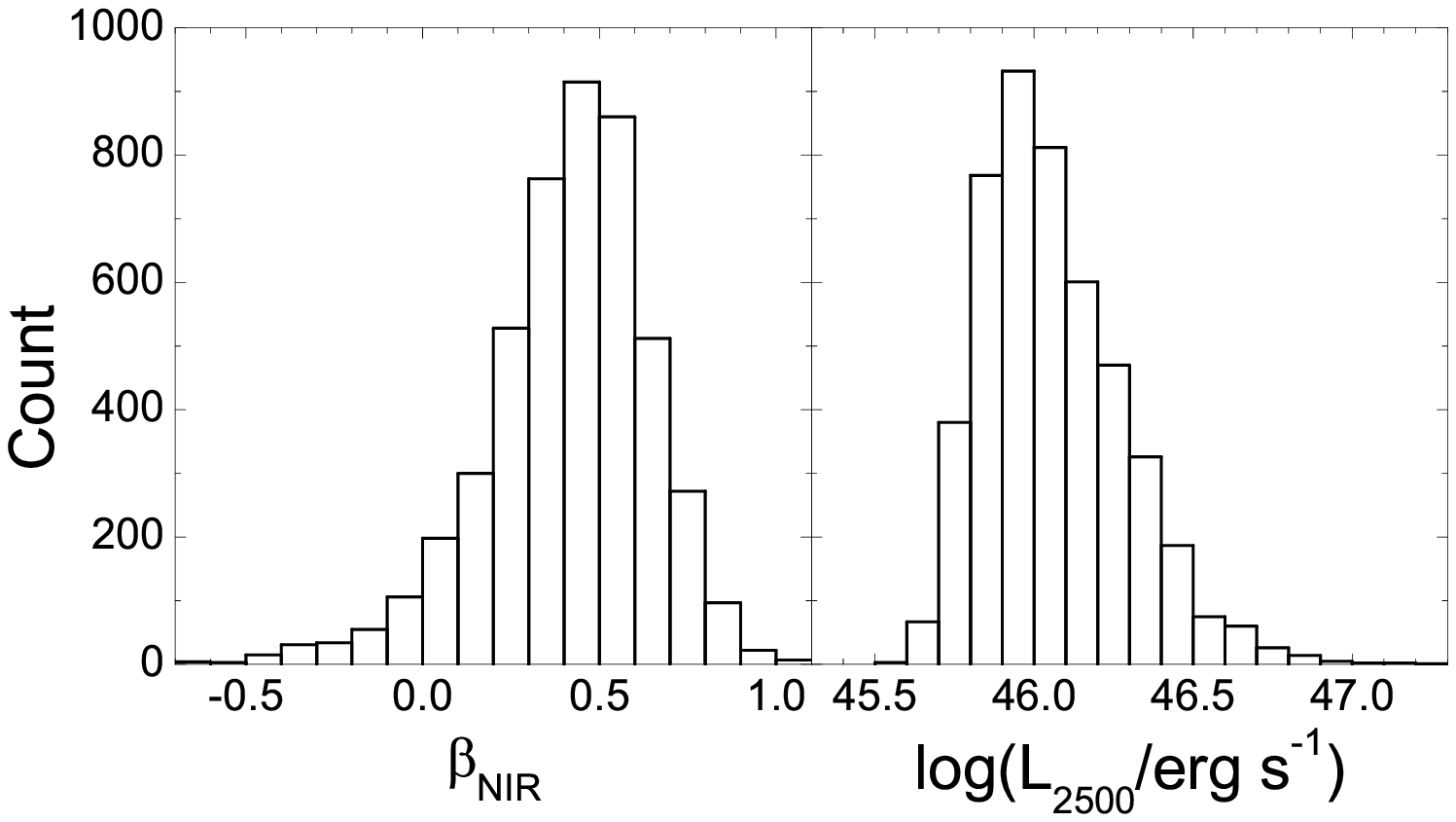}\caption{The number count distributions of the NIR (1-4$\mu\rm m$) slope $\beta_{\rm NIR}$ (left panel) and $\log(L_{2500})$ (right panel).}\label{fig1}
\end{figure}

We fit the NIR SEDs in the wavelength range of 1 to 4$\mu\rm m$. In this band, the emission from the hot dust at the maximum sublimation temperature begins to outshine the big blue bump (Elvis \etal\ 1994). And the host galaxy contribution to NIR emission is negligible for $z\sim2$ quasars (Hao \etal\ 2012). Therefore NIR emission is a good indicator of hot dust for our quasar sample. We present the distribution of the $\beta_{\rm NIR}$ parameter in Figure \ref{fig1}. The distribution peaks at 0.5, with a standard deviation of 0.24. Hao \etal\ (2011) found a average slope of 0.53 in a small SDSS quasar sample, in good agreement with ours. But the standard deviation of their sample is 0.37, apparently larger than ours. It might be ascribed to a much wider range of redshifts and luminosities in their sample.

\begin{table}
\caption{Correlation coefficients}
\label{tab_cor}
\begin{tabular}{lllllll}
\hline\hline
Sample & $\langle\log(l_{\rm E})\rangle$ & $L_{2500}\mid\beta_{\rm NIR}$  & $l_{\rm E}\mid\beta_{\rm NIR}$ & $M_{\rm BH}\mid\beta_{\rm NIR}$ &  $\beta_{\rm NIR}\mid$CF3 &  $\beta_{\rm NIR}\mid$CF1\\
\hline
Whole & - & -0.01(3$\rm e$-1) & 0.03(8$\rm e$-2) & -0.03(7$\rm e$-2) & 0.68($<1\rm e$-43) & -0.17($3\rm e$-33) \\
Sub1 & -0.87 & -0.01(9$\rm e$-1) & -- & -0.07(9$\rm e$-2) & -- & -- \\
Sub8 & 0.16 & -0.06(2$\rm e$-1) & -- & 0.05(3$\rm e$-1) & -- & -- \\
\hline\hline
Sample & BAI$\mid\beta_{\rm NIR}$ & FWHM$\mid\beta_{\rm NIR}$ & BAI$\mid$CF3 & FWHM$\mid$CF3 & BAI$\mid$CF1 & FWHM$\mid$CF1\\
\hline
Whole & 0.29($<1\rm e$-43) & 0.26($<1\rm e$-43) & 0.26($<1\rm e$-43) & 0.18(5$\rm e$-34) & 0.04(1$\rm e$-2)  & -0.03(5$\rm e$-2) \\
Sub1 & 0.15(4$\rm e$-4) & 0.05(2$\rm e$-1) & 0.10(1$\rm e$-2) & 0.03(4$\rm e$-1) & -0.03(5$\rm e$-1) & -0.01(7$\rm e$-1) \\
Sub8 & 0.55($<1\rm e$-43) & 0.49($3\rm e$-36) & 0.46(2$\rm e$-32) & 0.37(4$\rm e$-20) & 0.04(3$\rm e$-1) & -0.04($4\rm e$-1) \\
\hline
\end{tabular}
\tablecomments{For each entry, we list the Spearman rank correlation coefficient ($\rho_{\rm r}$) and the probability of the null hypothesis that the correlation is not present (P$_{\rm null}$). FWHM is FWHM(\civ) and $\langle\cdot\cdot\rangle$ denotes average.}
\end{table}

We first analyze the relationship between $\beta_{\rm NIR}$ and the quasar fundamental properties, such as $L_{2500}=\lambda L_{\lambda}(2500\A)$, $l_{\rm E}$ and $M_{\rm BH}$. In order to obtain a quantitative analysis, we the use Spearman correlation
coefficient ($\rho_{\rm r}$) to measure the statistical dependence between two quantities.
The coefficients are listed in Table \ref{tab_cor}, together with the probability of the null hypothesis that the correlation is not present (P$_{\rm null}$). As one can see, there is almost no dependence of $\beta_{\rm NIR}$ on the fundamental
parameters. It implies that the NIR slope is not directly regulated by the central engine. Recently, Krawczyk \etal\ (2013) found a dependence of mean NIR SEDs on luminosity. When looking at their figure 13 (the upper left panel) in detail, we however find that the dependence is very weak for quasars with $\log(L_{2500}/{\rm erg s^{-1}})\geq 45.41$. Since all of our quasars have $\log(L_{2500}/{\rm erg s^{-1}})\geq 45.5$ (Figure \ref{fig1}), our results don't conflict with theirs.

\begin{figure}
\epsscale{1}\plotone{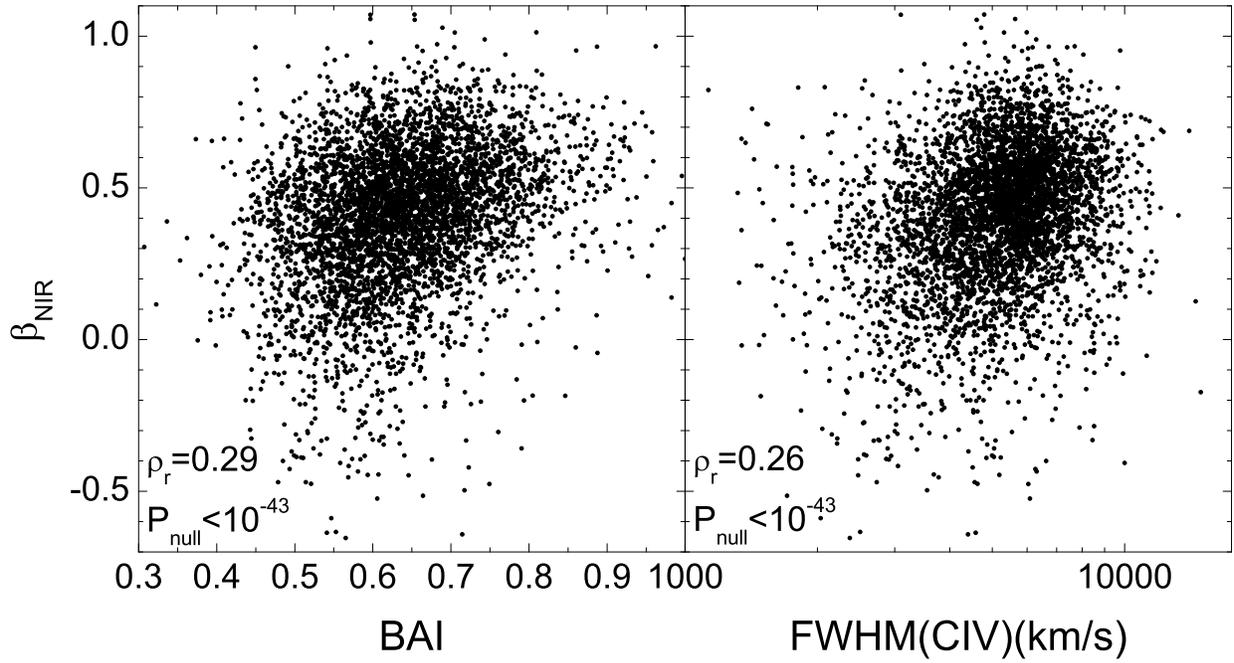}\caption{Left panel: $\beta_{\rm NIR}$ versus BAI, the blueshift and asymmetry index of \civ\ line.
Right panel: $\beta_{\rm NIR}$ versus FWHM(\civ). The results are for the whole sample. }\label{fig2}
\end{figure}

\begin{figure}
\epsscale{1}\plotone{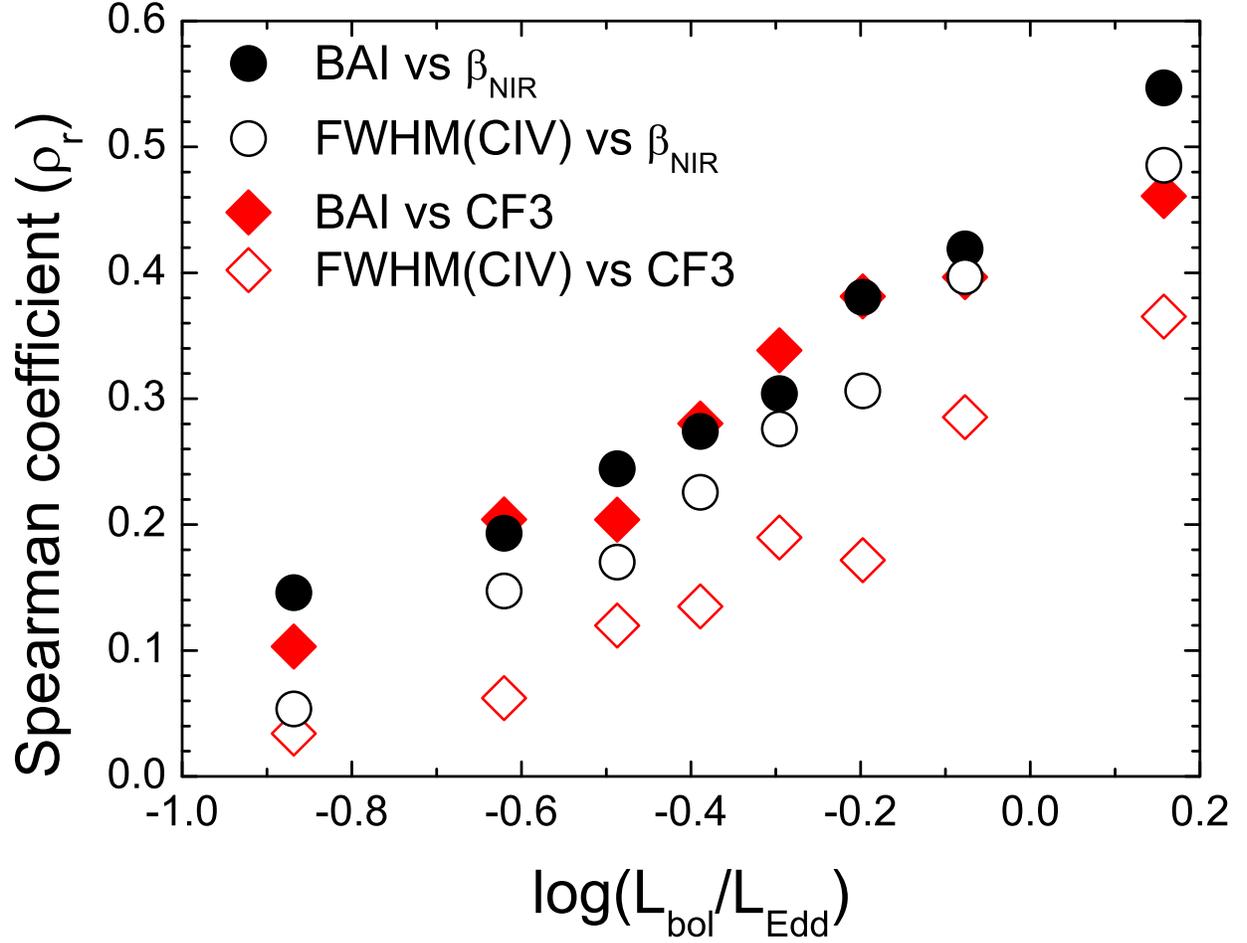}\caption{The Spearman correlation coefficients $\rho_{\rm r}$ for $\beta_{\rm NIR}$-BAI (filled black circle), $\beta_{\rm NIR}$-FWHM(\civ) (open black circle), CF3-BAI (filled red diamond) and CF3-FWHM(\civ) (open red diamond) as a function of the Eddington ratio.}\label{fig3}
\end{figure}

\begin{figure}
\epsscale{1}\plotone{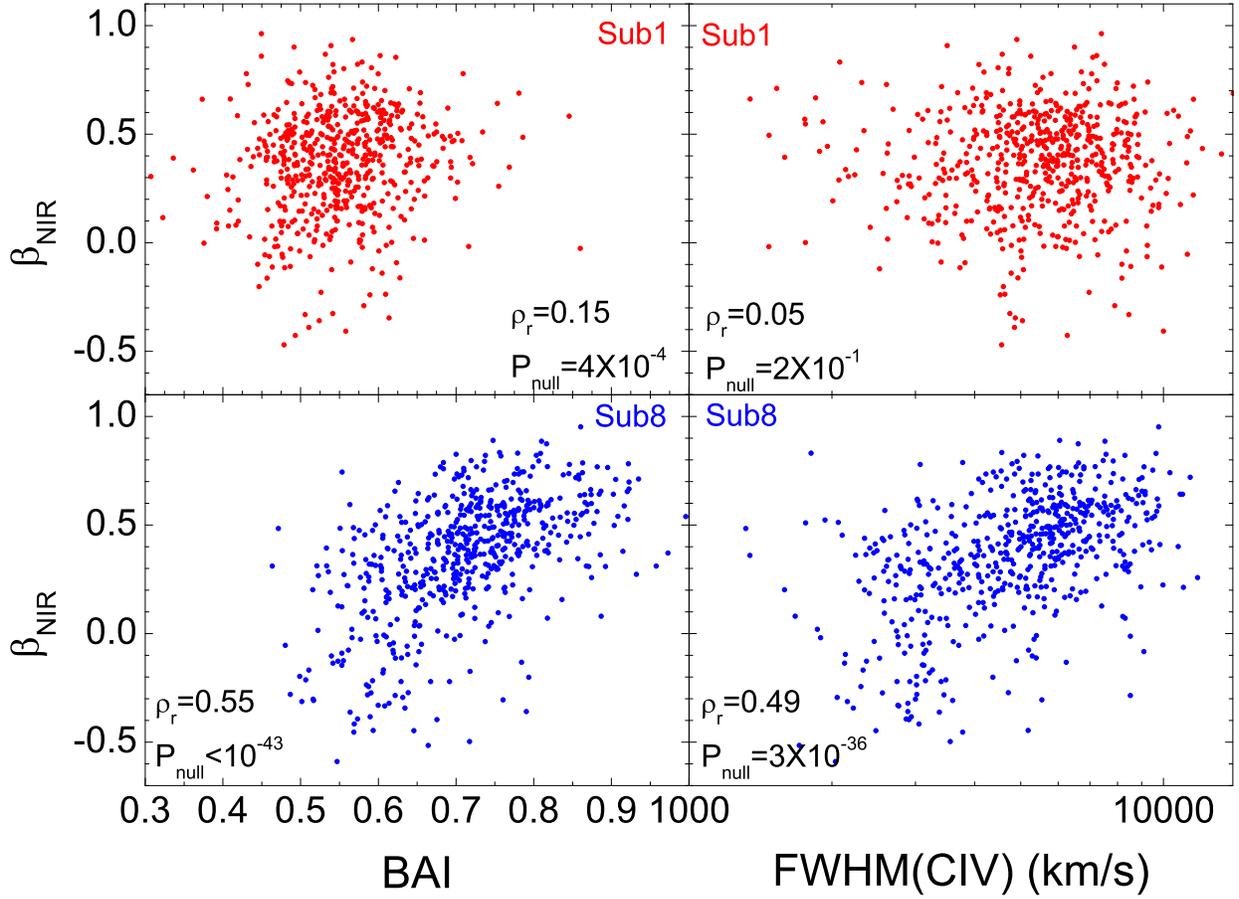}\caption{Left panels: $\beta_{\rm NIR}$ versus BAI; Right panels: $\beta_{\rm NIR}$ versus FWHM(\civ). The upper and lower panels show the results for the Sub1 and sub8 subsamples, respectively.}\label{fig4}
\end{figure}

Now we turn our attention to the outflow properties. The left panel of Figure \ref{fig2} shows $\beta_{\rm NIR}$ versus BAI. One can see a
clear trend of $\beta_{\rm NIR}$ increasing with BAI despite a large scatter. The Spearman test gives a moderate correlation coefficient of $\rho_{\rm r}=0.29$ (Table \ref{tab_cor}). Wang11 found a strong positive correlation between BAI and $l_{\rm E}$, which favors a radiatively driven outflow scenario. Through a comparative study of \civ\ and \mgii\ lines, Wang11 further found that the \civ\ lines in low $l_{\rm E}$ quasars are predominantly emitted by virialized gas (normal BEL gas) rather than outflows. Therefore, the dynamic range of outflows in these low $l_{\rm E}$ quasars is small, and the BAI measurement becomes insensitive to outflow properties. This effect may dilute the observed signal of the correlation between BAI and $\beta_{\rm NIR}$. In order to verify whether our analysis is valid, we divide the quasar sample into eight equally sized subsamples, named Sub$i$ ($i=1,2,..,8$), according to their $l_{\rm E}$. A small numerical suffix in the sample name denotes low $l_{\rm E}$. Each subsample is composed of about 591 quasars. We perform a Spearman test for each subsample and show the resultant $\rho_{\rm r}$ as a function of the subsample's mean $\log(l_{\rm E})$ in Figure \ref{fig3}. The correlation coefficient (see also Table \ref{tab_cor}) dramatically increases from 0.15 for Sub1 (nearly no correlation) to 0.55 for Sub8 (strong correlation). Figure \ref{fig4} show the scatter plots of $\beta_{\rm NIR}$ versus BAI for two extreme cases, Sub1 and Sub8. The Sub1 quasars tend to have \civ\ lines with small blueshift, indicative of negligible outflow contribution. Many among them even have a BAI of less than 0.5, which can not be ascribed to outflows, but is likely induced by the asymmetric profiles of normal BELs or inflows. The BAI for the Sub8 subsample is systematically higher and distributed in a wider range of 0.5 to 0.9, indicative of a outflow component dominating the \civ\ line. These results are consistent with our qualitative analysis. It is worth noting that similar differences between low and high $l_{\rm E}$ quasar populations are also detected in X-ray, radio and optical \feii\ emissions (Sulentic \etal\ 2000; Zamfir \etal\ 2008).

Another piece of evidence supporting our analysis can be found in the behavior of FWHM(\civ), which reflects the gas motion. For virialized BEL regions, FWHM(\civ) directly characterizes the virial velocity. For outflowing gases, FWHM(\civ) is a good indicator of outflow velocity. If the NIR emission really relates to outflows, FWHM(\civ) should behave similarly to the BAI. We present the relevant results in Figure \ref{fig2}, \ref{fig3} and \ref{fig4} and Table \ref{tab_cor}. The correlation between $\beta_{\rm NIR}$ and FWHM(\civ) is indeed very similar to that between $\beta_{\rm NIR}$ and the BAI, in the sense that the correlation is totally absent in Sub1 and $\rho_{\rm r}$ significantly increases with $l_{\rm E}$. We note that the correlation between FWHM(\civ) and the BAI itself is complicated (Wang11) and the corresponding $\rho_{\rm r}$ is only 0.22 in the entire sample. So the similarity between these two parameters is more likely induced by the same physical driver than any unknown secondary effect. Overall, these results all support our analysis that the contamination from normal BEL has  a nonnegligible impact at low $l_{\rm E}$. The correlations for the highest-$l_{\rm E}$ subsample are much stronger and tighter than those for low $l_{\rm E}$ ones, and very likely reflect the intrinsic connection between outflows and dust.

The physical meaning of $\beta_{\rm NIR}$ is not straightforward. It may characterize the ratio of hot dust to accretion disk emission. Alternatively it measures the dust temperature distribution if L$_1$ and L$_3$ are both dominated by dust emission. In the literature, many works adopted the infrared to UV luminosity ratio, which is readily interpreted as the relative amount of dust emission or the dust covering factor rather than the temperature distribution (e.g. Mor \& Trakhtenbrot 2011; Ma
\& Wang 2013). Here we use a parameter CF3=$\log(L_3/L_{2500})$ instead of $\beta_{\rm NIR}$ to repeat all of the above works. CF3 is very tightly correlated with $\beta_{\rm NIR}$ ($\rho_{\rm r}=0.68$, Table 1) and yields nearly identical results to $\beta_{\rm NIR}$. In the whole sample, CF3 is almost independent of $l_{\rm E}$ and $M_{\rm BH}$, and marginally dependent on $L_{2500}$ ($\rho_{\rm r}=-0.19$), consistent with Mor \& Trakhtenbrot (2011). The dependence of CF3 on the BAI is also of moderate strength ($\rho_{\rm r}=0.26$), while the dependence on FWHM(\civ) is weak ($\rho_{\rm r}=0.18$). Most interestingly, these two dependencies also strengthen with increasing Eddington ratio (Figure \ref{fig3} and Table \ref{tab_cor}). Because of the strong correlation and the same manner between CF3 and $\beta_{\rm NIR}$, these two quantities most likely represent the same factor, i.e. the hot dust emission relative to the disk emission. Compared to CF3, $\beta_{\rm NIR}$ is less affected by variation, inclination effect and dust extinction. This might be the reason why the correlations of outflow properties with $\beta_{\rm NIR}$ are slightly stronger than those with CF3.

Finally we investigate another parameter, CF1=$\log(L_1/L_{2500})$. CF1 is moderately dependent on luminosity ($\rho_{\rm r}=-0.27$), weakly related to $\beta_{\rm NIR}$ ($\rho_{\rm r}=-0.17$, Table 1), and independent of $l_{\rm E}$ and $M_{\rm BH}$. Most importantly, CF1 is totally uncorrelated with the BAI and FWHM(\civ) in the entire sample and each subsample (Table. \ref{tab_cor}). It indicates a different origin of L$_1$ ($\sim1\mu$m) from L$_3$ ($\sim4\mu$m) which is consistent with previous studies. Elvis \etal\ (1994) suggested that dust emission rises above disk emission around 1$\mu$m. Recently, Mor \& Trakhtenbrot (2011) decomposed the NIR SEDs of SDSS quasars and found that hot dust emission only dominates the SED between 2$\mu$m and 5$\mu$m.

\section{Summary and Discussion}\label{secsum}

Taking advantage of the WISE and the SDSS data releases, we construct a large sample of z$\sim2$ quasars to study the correlation
between the outflow properties and hot dust emission. We measure the NIR SED slope ($\beta_{\rm NIR}$), in the rest-frame wavelength of 1$\mu\rm m$ to 4$\mu\rm m$, to quantify the relative amount of hot dust. The NIR slope is independent of the quasar fundamental parameters, such as luminosity, Eddington ratio and black hole mass for our luminous quasars. We find interesting and moderate dependencies of $\beta_{\rm NIR}$ on both the BAI and FWHM(\civ) in the whole sample. Most interestingly, these two correlations dramatically strengthen with Eddington ratio. Note that using CF3=$\log(L_3/L_{2500})$ instead of $\beta_{\rm NIR}$ leads to the same results. In high Eddington ratio quasars, \civ\ lines are supposed to be dominated by outflow component so that the BAI and FWHM(\civ) measure well the general property and velocity of outflows, and thus strongly correlate with hot dust emission. While in low Eddington ratio quasars, \civ\ lines are predominantly emitted by virialized gas so the BAI and FWHM(\civ) can not reliably reflect the outflow properties. These low Eddington ratio quasars therefore dilute the correlations. We suggest that the correlations observed in high Eddington ratio quasars very likely reflect the true dependence of the hot dust emission on the outflow properties. As BAI increases from 0.5 to 0.9 or FWHM(\civ) increases from 2000 to 10000km/s, $\beta_{\rm NIR}$ increases on average from about zero to 0.6. This means that the 4$\mu\rm m$ emission relative to 1$\mu\rm m$ for a large BAI/FWHM(\civ) quasar is roughly twice as large as that for a small BAI/FWHM(\civ) quasar.

Other factors might also produce the observed correlations. For a geometrically thin disk, disk emission roughly scales with the cosine of the disk inclination angle with respect to the line of sight. Both $\beta_{\rm NIR}$ and CF3 measure the ratio of hot dust to disk emission, so these two parameters will increase with increasing inclination angle. In an equatorial wind model (Murray \etal\ 1995), the larger is the inclination angle, the smaller is the BAI and the larger is FWHM(\civ). Opposite correlations of the BAI with hot dust emission and FWHM(\civ) to ours are expected. In a polar outflow case (Zhou \etal\ 2006), both the BAI and FWHM(\civ) increase as the inclination angle decreases. Again, the inclination effect fails to produce the observed dependencies of hot dust emission on the BAI and FWHM(\civ). Moreover, the correlations with CF3 may possibly be produced by an inclination-dependent extinction, since the UV band is more sensitive to dust extinction than the NIR band. However this hypothesis can be ruled out by the absence of correlations with CF1$=\log(L_1/L_{2500})$. Altogether the inclination effect seems unlikely to produce the observational results, on the contrary, it is one of the major sources of scatter in these correlations. It is also possible that outflows have nothing to do with dust and the correlations are induced by a third factor that simultaneously governs outflows and dust emission. Although we can not rule out this possibility, it seems unlikely since $\beta_{\rm NIR}$ and CF3 are almost independent of the fundamental parameters.

In a forthcoming paper, we perform similar studies using BAL quasars (S. Zhang \etal\ 2013 in preparation) and find significant correlations of $\beta_{\rm NIR}$ with the velocities and absorption strength of BALs, in good agreement with the results presented here. One important question arises as to how the outflows relate to the hot dust in quasars. We propose two possible mechanisms. The first one is that outflows originally contain dust or dust is manufactured in outflows (Elvis \etal\ 2002). The dust carried by outflows is heated by the central engine to emit at the sublimation temperature, meanwhile dust absorption contributes to the outflow acceleration. The reverberation mapping results suggested that hot dust is located at the outer boundary of BEL regions (Suganuma \etal\ 2006). The outflow launching region is also suggested to be co-spatial with or outside of the BEL regions. This coincidence implies the possibility of dusty outflows. In this scenario, quasars become an important source of cosmic dust, which may affect our understanding of several important issues in astrophysics and cosmology (see Elvis \etal\ 2002). The second mechanism is that outflows interact with the surrounding dusty torus or interstellar medium (ISM) and strip off dusty gas from these cold and dense clouds. The dust is then heated up by the central UV source to radiate in the NIR band. This idea is supported by hydrodynamical simulation in which outflows can break dense ISM clouds into warmer diffuse filaments (Wagner et al. 2013). If this mechanism works, a combined study of UV and NIR data will provide an avenue for understanding how the mechanical feedback proceeds.

\acknowledgements We thank the anonymous referee for a helpful report. This work is supported by NSFC (11073017, 11033007), 973
program (2009CB824800), NCET-11-0879 and the Fundamental Research
Funds for the Central Universities. Funding for the SDSS and SDSS-II has been provided by the Alfred P. Sloan
Foundation, the Participating Institutions, the National Science
Foundation, the U.S. Department of Energy, the National
Aeronautics and Space Administration, the Japanese Monbukagakusho,
the Max Planck Society, and the Higher Education Funding Council
for England. The SDSS Web site is http://www.sdss.org/.

\end{document}